Title: Two-electron F′ centers in alkali halides: a saddle point approach. I. General and semicontinuum analyses
Author: Mladen Georgiev (Institute of Solid State Physics, Bulgarian Academy of Sciences, 1784 Sofia, Bulgaria)
Comments: 14 pages including 2 figures and 1 table, all pdf format
Subj-class: cond-mat

The F′ center in an alkali halide forms when an anion vacancy traps two electrons which is the prerequisite of a diatomic molecule. Indeed, the center may displace left or right along <110> in a (110) plane, due to its coupling to the $B_{1u}$ vibrational mode of < 110 > polarization, changing its position from [000] to [$\underline{1}$10] or to [1$\underline{0}$0], respectively, These are merely examples for the migrational steps of the anion vacancy. On jumping from the initial position to the final position the F′ center passes through a saddle-point which configuration is molecule-like being conformed by two neighboring semi-vacancies along < 110 >. Each semi-vacancy traps one electron to change its effective charge from +½ to −½ forming a semi- F′-center. We outline the basic theory so as to perform calculations of the F′ eigenenergies in two configurations, vacancy-centered and saddle-point, to see if the former energy can surpass the latter so as to make the saddle-point configuration more favorable energetically.

1. Introduction

The F' center (electronically, two electrons trapped at an anion vacancy) is interesting in that it provides an example for a localized electronic dimer with possible implications for fields where electron dimers are important, e.g. color center physics and high-temperature superconductivity. (See Ref. [1] for a review of the F' work prior to 1988.) Indeed, it is highly desirable to learn just how an electron dimer forms at the vacancy site and in that respect the role played by lattice polarization for keeping the two trapped electrons together despite their Coulomb repulsion.

An important post 1988 work has deduced excited singlet and triplet states of the F' center without laying any special emphasis on the lattice polarization.[2] Nevertheless, a more recent study has confirmed that lattice polarization does play a particular role for keeping the electrons bound via the negative-U effect.[3] This is an on-site mechanism of enhancing considerably the (di-) electron trapping energy at a given site through its quadratic dependence on the electron–vibrational mode coupling strength.[4]

The traditional negative-U mechanism is operative in the equilibrium configuration of the trapped electron site.[4] But the F' center may also displace along <110> in a (100) plane which is conceivable despite the hindering effect of its trapped electrons in ground state.[5] Now, the question is raised as to the importance of parallel intersite effects, e.g. the ones arising at the saddle point between two neighboring equilibrium configurations.[5] At the saddle point the original F' center site splits into two semivacancies, each one hosting one of the F' electrons at a time. This is illustrated in Figure 1 similar to the $F_A'$ centers.[5] The $F_A'$ center is an F' center nearest-neighboring a smaller size impurity ion, such as $Na^+$ or $Li^+$ in KCl in a (110) plane. To make the picture authentic to the F' center we should only substitute $K^+$ for the impurity $Na^+$ ion in accommodating to Luty's drawing.

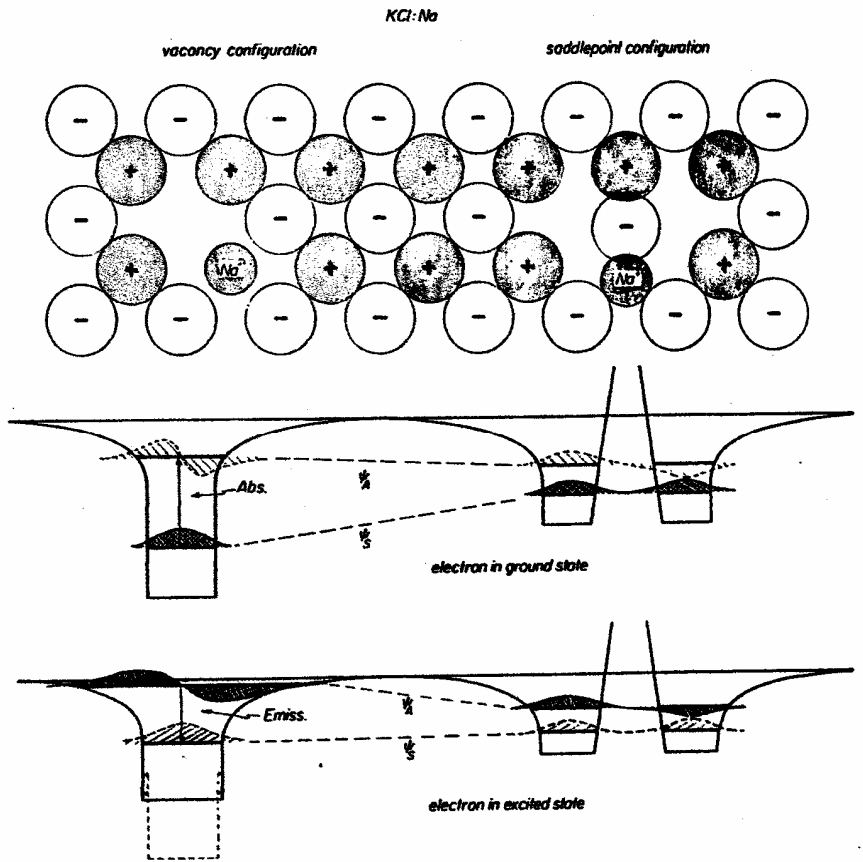

Figure 1. (110) plane of an impure KCl:Na crystal containing $F_A$(Na) centers, from Reference [5]. Full- vacancy configuration (left), semi-vacancies (saddle point) configuration (right). The semi-continuum electronic potentials pertaining to the two configurations are also shown below along with the corresponding densities of the two lowest electronic states. Turning to the F center in a pure crystal would require substituting a normal $K^+$ ion for the $Na^+$ impurity.

For the forthcoming analysis it might be useful to derive analytic expressions for the electronic potentials in the two vacancy configurations of Figure 1. We ultimately aim at estimating the electron-mode coupling constant pertinent to the saddle-point problem and assess its feasibility. Figure 1 suggests that as the <110> halogen ion starts displacing towards the saddle point, the original left-hand cavity squeezes and a cavity to the right of the impurity opens up and grows at the expense of the former cavity. The process passing through the saddle point, it ends up with the displacement of the anion vacancy to a neighbouring position to the right of the original anion vacancy. We see that the semi-vacancy configuration is an ingredient step of the anion vacancy migration.

Each vacancy potential is seen composed of a flat bottom part inside the cavity followed by a small Coulomb tail just outside.[1] We define $V_\pm$, the semi-vacancy potentials, as

$V_- (r_0, Q) = V_M (r_0) r_0 / (Q + \Delta_0)$

$V_+ (r_0, Q) = V_M (r_0) r_0 / (r_0 - Q + \Delta_0)$,

where Q ($0 \leq Q \leq r_0$) is the <110> displacement of the halogen ion to the left as in Figure 1. Subsequently the overall potential along the Q-axis is:

$V(r_0, Q) = V_- (r_0, Q) \theta(Q) \theta(Q + \Delta_0) + V_+(r_0, Q) \theta(Q) \theta(r_0 - Q + \Delta_0)$

Here $\Delta_0 > 0$ is a small residual bottom size when the site is fully occupied by a normal lattice ion. Introducing $\Delta_0$ keeps the incipient cavity potential from being singular.

It is easy to verify the distribution of $V_\pm$ on both sides of the splitting $Q = \frac{1}{2} r_0$ line. The separation between $V_\pm$ along Q is maximal ($r_0$) when $Q = \frac{1}{2} r_0$. The minimum (0) obtains at $Q = 0, r_0$ as the spitting between the semi-vacancies disappears to produce a single full vacancy to the left or to the right of the impurity, respectively. Accordingly, $Q = 0, r_0$ corresponds to a full-vacancy configuration and $Q = \frac{1}{2} r_0$ to a semivacancy configuration. Here and throughout $r_0$ is the full-vacancy cavity radius, $V_M$ is Madelung's potential at the vacancy site, and $\theta(x)$ is the step function ($\theta(x) = 1, x \geq 0, \theta(x) < 0, x < 0$).

We further regard Q as the <110> coupled-mode coordinate ($-r_0 \leq Q \leq +r_0$). (Extending Q to negative values is straightforward.) Indeed, the mode coordinate couples to and splits the full-vacancy potential into two semi-vacancy potentials. Using the above analytic expressions we derive an electron-mode coupling constant by differentiating in Q and taking the derivative at Q=0:

$G(r_0) = \langle \phi_1 | [\partial V_- (r_0, Q) / \partial Q] \theta(Q) \theta(Q + \Delta_0) + V_- (r_0, Q) \theta(Q) \delta(Q + \Delta_0) +$

$\qquad V_- (r_0, Q) \theta(Q + \Delta_0) \delta(Q) +$

$\qquad [\partial V_+ (r_0, Q) / \partial Q] \theta(Q) \theta(r_0 - Q) - V_+ (r_0, Q) \theta(Q) \delta(r_0 - Q + \Delta_0) +$

$\qquad V_+ (r_0, Q) \theta(r_0 - Q + \Delta_0) \delta(Q) |_{Q=0} | \phi_2 \rangle$

$= V_M (r_0) r_0 \langle \phi_1 | \{-[1/(Q+\Delta_0)^2] \theta(Q+\Delta_0) + [1/(r_0-Q+\Delta_0)^2] \theta(r_0-Q+\Delta_0)\} \theta(Q) |_{Q=0} | \phi_2 \rangle$

$$\sim -V_M(r_0)\, r_0 \langle\phi_1|\,[(1/\Delta_0)^2 - (1/r_0)^2]\,|\phi_2\rangle$$

We thus arrive at a simple coupling constant for the pure crystalline material. However, G is growing larger if a smaller-size impurity is substituted for a vacancy neighboring host cation, apparently by opening more space for the displacing particle.

Figure 1 suggests applying molecular methods to describe the saddle point F' center. As a scope of the present study, we consider it interesting to see whether under certain conditions the saddle point configuration is not more favorable energetically than the equilibrium configuration. In other words, what we want to know is whether the F' electron dimer is overwhelmingly on-site or there are inter-site exceptions too.

## 2. F' Hamiltonian

At any lattice configuration, the F' Hamiltonian is composed of an electron term under $H_{el}$, a lattice term under $H_{latt} = H_{latt}' + \frac{1}{2}M\omega^2 Q^2$, and their interaction energy term $\propto GQ$:

$$H = H_{el} + H_{latt} + H_{int} \equiv H_{el} + H_{latt}' + \tfrac{1}{2}M\omega^2 Q^2 + 2GQ \qquad (1)$$

Here M is the mass of the halogen ion vibrating along $<110>$, $\omega$ is the vibrational frequency, Q is the respective vibrational mode (configurational) coordinate, G is the electron-mode coupling constant. The electron-mode coupling energy is linear in the configurational coordinate, the factor 2 arising from the coupling assumed the same for each of the two F' center electrons. To 1st-order perturbation, the Q-dependent terms in (1), $\frac{1}{2}M\omega^2 Q^2 + 2GQ$, generate adiabatically an electronic eigenenergy of the form

$$E_\pm(Q) = \tfrac{1}{2} M\omega^2 Q^2 \pm \tfrac{1}{2}\sqrt{[16 G^2 Q^2 + V_{12}^2]} \qquad (2)$$

in which $V_{12} = \langle 1|\, H_{int} + e^2/\kappa r_{12}\,|2\rangle$ is the coupling energy of $|1\rangle$ and $|2\rangle$ at the $Q = Q_C$ saddle-point configuration coordinate and $r_{12} = |r_2 - r_1|$. In the two-site problem as in Figure 1 the electronic states $|1\rangle$ and $|2\rangle$ are eigenstates of $H_{el}$, e.g. ground s-like states or excited s- or p-like states, centered at neighboring equilibrium sites. These position states (static basis) mix up to produce a resonance splitting $V_{12}$ of the adiabatic energy at the $Q_C$ crossover. The other term in $V_{12}$ is the e-e correlation energy.

More specifically, the electronic Hamiltonian $H_{el}$ is composed of kinetic and potential energy terms for each of the F' center electrons plus their correlation energy:

$$H_{el} = \sum_{i=1,2} [p_i^2/2m_i + V_i(r_i)] + e^2/\kappa|r_2 - r_1| \qquad (3)$$

For comparative though not very accurate calculations the electronic potential $V_i(r_i)$ may be taken in the semicontinuum form composed of a constant potential at close range (inside the cavity) followed by a Coulomb tail at long range (see Section 5).[1]

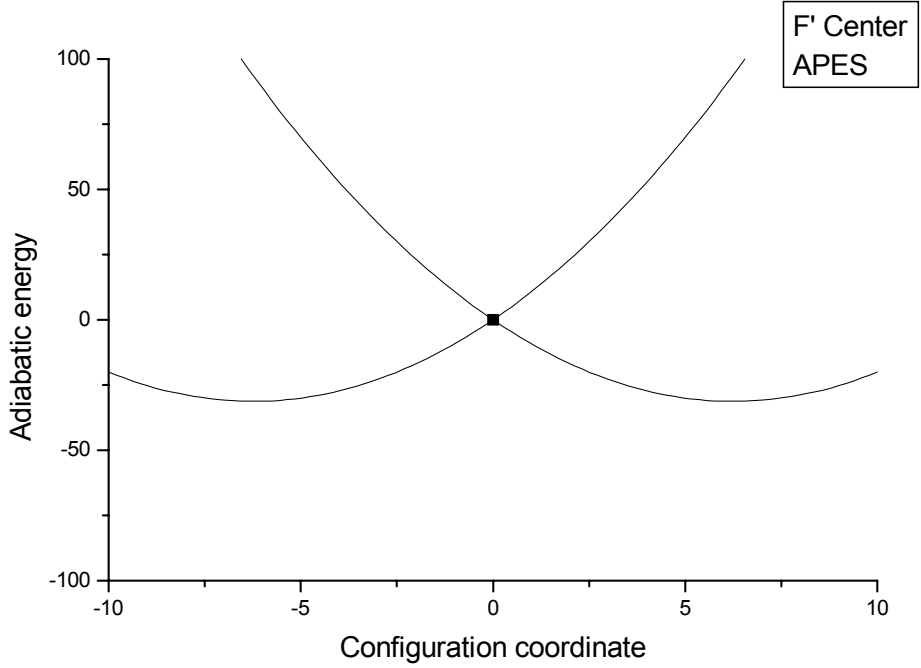

Figure 2. Schematic configurational coordinate diagram along < 110 > of the F′ center following eqn. (2). The parameters used for the calculation are: $K = M\omega^2 = 1.6$ eV/Å$^2$, $G = 5$ eV/Å, $V_{12} = 0.2$ eV. The black square at the origin marks the position and size of the resonant splitting of the electronic energy at the crossover coordinate.

The configurational-coordinate dependence of the eigenenergy $E_\pm(Q)$ is depicted in Figure 2 (schematically) at strong coupling $2G^2/K > V_{12}$. Here $K = M\omega^2$ is the stiffness. The two lateral wells, left and right, occuring at

$$Q_{0\pm} = \pm\sqrt{[64G^4 - V_{12}^2 K^2]} / 4GK \qquad (4)$$

relate to the two physically identical equilibrium configurations of the on-site F' center in Figure 1, the interwell barrier is at the saddle-point (crossover) configuration in-between. We remind that the electronic potential, single-well laterally, splits into two semi-wells at saddle point, as shown in Figure 1.

To adiabatic approximation separating the electronic and ionic coordinates, the two-branch adiabatic electron energy (2) serves as potential energy for the vibronic mode coupled to the F' center. This is the eigenvalue problem of a double-well oscillator which has not been solved rigorously. Near the well bottoms, the eigenstates are approximated for by harmonic oscillator eigenfunctions at a renormalized frequency dependent on the coupling strength:

$$\omega_{renI} = \omega \sqrt{[1 - (V_{12}^2/4E_{CE})^2]} \tag{5}$$

where $E_{CE} = 2G^2/K$ is the coupling energy. The complete adiabatic wavefunction of the coupled F' center is a sum of products of electronic and vibronic eigenstates.

### 3. Tunneling splitting at the saddle point configuration

For dealing with the eigenvalue problem of (1) it would be profitable to look at the earlier work though from a different angle. Luty refers to an unpublished work by Strozier in which he considers the energy of the saddle-point F center.[5] We will follow his arguments even though our F' problem is more plentiful having one extra electron. Introducing Slater's orbitals

$$\psi(r) = (\alpha^3/2\pi) \exp(-\alpha r) \tag{6}$$

Strozier defines the symmetric and antisymmmetric combinations

$$\psi_S = \psi(r_1) + \psi(r_2)$$

$$\psi_A = \psi(r_1) - \psi(r_2)$$

whose eigenenergies are

$$E_S = \langle\psi_S|H|\psi_S\rangle / \langle\psi_S|\psi_S\rangle$$
$$E_A = \langle\psi_A|H|\psi_A\rangle / \langle\psi_A|\psi_A\rangle \tag{7}$$

Minimizing he gets

$$E_S = [\langle\psi_1|H|\psi_1\rangle + \langle\psi_1|H|\psi_2\rangle] / [1 + \langle\psi_1|\psi_2\rangle]$$
$$E_A = [\langle\psi_1|H|\psi_1\rangle - \langle\psi_1|H|\psi_2\rangle] / [1 - \langle\psi_1|\psi_2\rangle] \tag{8}$$

It should be mentioned that the energy difference $\Delta E_{AS} = E_A - E_S$ is the ground state tunneling splitting at saddle point. Strozier's electronic object has the same symmetry as ours. Its effective charge is 0, ours is $-1$ (equilibrium configuration) which distributes as $-\tfrac{1}{2}$ to each of the semi-vacancies (saddle point configuration). (Each semi-vacancy carries an effective charge of $+\tfrac{1}{2}$). All the effective charges are in elementary electronic charge units. We sometimes proceed accordingly by adopting the notion of *F' electron*, an often used simplification against the neutral vacancy (F center) background. In some cases this helps reduce the complex two-electron problem to one of a single electron.

Inserting (1) in (8) we obtain

$$E_S = (E_1 + \langle 1| e^2/\kappa|\mathbf{r}_2 - \mathbf{r}_1| + H_{latt} + H_{int}|1\rangle + V_{12}) / (1 + S_{12})$$

$$E_A = (E_1 + \langle 1| e^2/\kappa|\mathbf{r}_2 - \mathbf{r}_1| + H_{latt} + H_{int} |1\rangle - V_{12}) / (1 - S_{12}) \qquad (9)$$

where $V_{12}$ has been defined in Section 2, while $S_{ik}$ are the overlap integrals. At $S \equiv S_{12} \ll 1$, the tunneling splitting is found to be

$$\Delta E_{AS} = [(E_1 + \langle 1| e^2/\kappa|\mathbf{r}_2 - \mathbf{r}_1| + H_{latt} + H_{int} |1\rangle - V_{12})(1 + S_{12}) -$$

$$(E_1 + \langle 1| e^2/\kappa|\mathbf{r}_2 - \mathbf{r}_1| + H_{latt} + H_{int} |1\rangle + V_{12})(1 - S_{12})] / (1 - S_{12}^2)$$

$$= 2[-V_{12} + (E_1 + \langle 1| e^2/\kappa|\mathbf{r}_2 - \mathbf{r}_1| + H_{latt} + H_{int} |1\rangle)S] \qquad (10)$$

as it should, for it reduces to $|\Delta E_{AS}| \sim 2|V_{12}|$ at small wave function overlap.

The saddle-point barrier as measured from the well bottom reads:[6]

$$E_{BI} = E_{CE} [1 - (V_{12} / 4E_{CE})]^2 \qquad (11)$$

It is small at weak coupling ($V_{12} \leq 4E_{CE}$) and large at strong coupling ($V_{12} \ll 4E_{CE}$). In the former case (i) the F' center is smeared around the barrier as it is shared almost equally between the saddle point and the equilibrium sites. In the latter case (ii) the F' center will be smeared over the equilibrium sites as little of its density will go to the barrier region.

For a quantitative criterion, we make use of the reorientational barrier of off-center ions in alkali halides:[7]

$$E_{BII} = (1/8)I\omega_{renII}^2 = 2[(D_b - D_c)/G](E_{CE}^2/K)[1 - (V_{12}/4E_{CE})^2]^2 \qquad (12)$$

Here $D_b$ and $D_c$ are third-order-tensor electron-mode coupling constants.[7] Comparing it with the saddle-point barrier will show just how the F' molecule smears along a given orientation: We take the ratio of (12) to (11) to get

$$E_{BII} / E_{BI} = 2(E_{CE} / K)[(D_b - D_c) / G][1 + (V_{12} / 4E_{CE})]^2 \qquad (13)$$

It will be seen that case (i) arises at $E_{BII} / E_{BI} \gg 1$, while $E_{BII} / E_{BI} \ll 1$ applies to case (ii). Indeed, with nearly all the parameters (G, K and D, in particular) having standard values, the ratio (13) is mainly controlled by $V_{12}$ through the factor $[1 + (V_{12} / 4E_{CE})]^2$ which is numerically within the limits $1 \div 4$ for coupling from strong to weak.

The occurrence of a tunneling splitting (10) due to the exchange of electrons between the semi-vacancies gives rise to a chemical bond within the divacancy at saddle point.

## 4. F' molecule

There are different ways of arranging a structure of two semi-vacancies attached to a displaced anion at a saddle-point site. The three form something like a molecule of a given symmetry. Both linear and triangular configurations are conceivable though we shall presently focus on the former. Each anion having 12 nearest neighbor anion sites, there are 12 potential anion-anion vacancy pairs, there should be a corresponding number of F' molecule orientations which are quantized. Accordingly, each F' molecule can perform classical jumps or tunneling rotation between the reorientational sites leading to the occurrence of rotational energy bands.[7] As above, comparing the reorientational barrier (13) with the saddle-point barrier (11) will show how the F' molecule smears along a given orientation.

The eigenstates and eigenenergies of an F' molecule, such as that composed of two semi-vacancies separated at $d_0$ can be worked out by analogy with diatomic molecules.[8] In a related study the vacancy wave functions have been obtained in the semicontinuum approximation which gives them as spherical Bessel functions $j_l(Kr)$ inside the cavity and Coulomb functions $R_{nl}(\alpha r)$ at longer range.[1] (However, account should be taken of the electronic correlations too.) One way or the other, if both constituents are in their ground electronic 1s-state, then for a molecular $^1S$ ground singlet

$$\Psi_{1S}(\mathbf{r}_1,\mathbf{r}_2) = N \, [\psi_{1s}(\mathbf{r}_1)_L \psi_{1s}(\mathbf{r}_2)_R + \psi_{1s}(\mathbf{r}_2)_L \psi_{1s}(\mathbf{r}_1)_R] \qquad (14)$$

where the spin states of the constituents are clearly indicated, while the suffixes L,R stand for left-hand semi-vacancy and right-hand semi-vacancy in the drawing of the saddle point configuration in Figure 1. If the reference for the electronic coordinates is set at the halogen ion splitting the two semi-vacancies, then $r_1 = |-½ \mathbf{d}_0 + \mathbf{u}_1|$, $r_2 = |+½ \mathbf{d}_0 + \mathbf{u}_2|$. We remind that $\mathbf{d}_0$ is the vacancy splitting vector from left to right. Next, if the electron at one of the semi-vacancies is in the excited 2p-state, then for a molecular $^1P$ excited singlet

$$\Psi_{1P}(\mathbf{r}_1,\mathbf{r}_2) = N \, [\psi_{1s}(\mathbf{r}_1)_L \psi_{2p}(\mathbf{r}_2)_R + \psi_{1s}(\mathbf{r}_2)_L \psi_{2p}(\mathbf{r}_1)_R] \qquad (15)$$

Clearly, the vacancy-centered equilibrium configuration obtains at $d_0 = 0$.

The non-correlated semi-continuum radial wave functions are:

$\psi(r)_{1s} = A_{k0} \, j_0(Kr) = A_{k0} \, (Kr)^{-1} \sin(Kr)$ ( in )

$\psi(r)_{1s} = B_{10} \, R_{10}(\alpha r) = B_{10} \exp(-\alpha r)$ (out)

$\psi(r)_{2p} = A_{k1} \, j_1(Kr) = A_{k1} \, (Kr)^{-1} [\, \sin(Kr)/(Kr) - \cos(Kr) \,]$ (in)

$\psi(r)_{2p} = B_{21} \, R_{21}(\beta r) = B_{21} \, (\beta^3/\pi)^{½} \, r \exp(-\beta r)$ (out), $\qquad (16)$

etc. The angular components are available from textbooks. The e-e correlations are most easily accounted for in ground state. A multiplicative factor of the form

$$\psi_{1S}(\mathbf{r}_1,\mathbf{r}_2) = C\exp(-\alpha r_{12}) \qquad (17)$$

has often been applied to where $r_{12} = |\mathbf{r}_2 - \mathbf{r}_1|$ along with simpler e-e correlation terms.[1] Following that, the functional

$$E_{F'} = \int \Psi_{1S}(\mathbf{r}_1,\mathbf{r}_2)\psi_{1S}(\mathbf{r}_1,\mathbf{r}_2) H_{F'} \Psi_{1S}(\mathbf{r}_1,\mathbf{r}_2)\psi_{1S}(\mathbf{r}_1,\mathbf{r}_2) \, d\mathbf{r}_1 d\mathbf{r}_2 \qquad (18)$$

is minimized in $\alpha$ to calculate the ground state energy.

The energy of the vacancy-centered configuration in a few crystalline hosts has been computed earlier using Wang's polaron theory.[1,8] For that matter, we shall presently focus on the saddle-point configuration under similar conditions. Starting with the non-correlated wave functions for simplicity, we get from (14) and (18):

$$E_{F'} = <r_1, L\,|<r_2, R\,|\,H_{F'}\,|\,r_1, L>|\,r_2, R> + <r_1, L\,|<r_2, R\,|\,H_{F'}\,|\,r_2, L>|\,r_1, R>$$

$$+ <r_2, L\,|<r_1, R\,|\,H_{F'}\,|\,r_1, L>|\,r_2, R> + <r_2, L\,|<r_1, R\,|\,H_{F'}\,|\,r_2, L>|\,r_1, R>$$

$$= E_{11LL}\,S_{22RR} + E_{22RR}\,S_{11LL} + E_{11LR}\,S_{22RL} + E_{22RL}\,S_{11LR} +$$

$$+ E_{11RL}\,S_{22LR} + E_{22LR}\,S_{11RL} + E_{11RR}\,S_{22LL} + E_{22LL}\,S_{11RR} +$$

$$<r_1, L\,|<r_2, R\,|\,H_{F'12}\,|\,r_1, L>|\,r_2, R> + <r_1, L\,|<r_2, R\,|\,H_{F'12}\,|\,r_2, L>|\,r_1, R> +$$

$$<r_2, L\,|<r_1, R\,|\,H_{F'12}\,|\,r_1, L>|\,r_2, R> + <r_2, L\,|<r_1, R\,|\,H_{F'12}\,|\,r_2, L>|\,r_1, R> \qquad (19)$$

In deriving (19) use has been made of

$$H_{F'} = H_{F'1} + H_{F'2} + H_{F'12}$$

according to (3). Here

$$E_{iiXY} = <r_i, X\,|\,H_{F'i}\,|\,r_i, Y> \qquad (20)$$

$$S_{iiXY} = <r_i, X\,|\,r_i, Y> \qquad (21)$$

for $i = 1, 2$ and $X, Y = L, R$ are the energy and overlap terms, respectively. Assuming wave functions localized strongly within the semi-vacancies, the off-diagonal terms are small and will be neglected. Under these conditions we get

$$E_{F'saddle} \approx E_{11LL}\,S_{22RR} + E_{22RR}\,S_{11LL} + E_{11RR}\,S_{22LL} + E_{22LL}\,S_{11RR} +$$

$$+ <r_1, L\,|<r_2, R\,|\,H_{F'12}\,|\,r_1, L>|\,r_2, R> + <r_1, L\,|<r_2, R\,|\,H_{F'12}\,|\,r_2, L>|\,r_1, R> +$$

$$<r_2, L\,|<r_1, R\,|\,H_{F'12}\,|\,r_1, L>|\,r_2, R> + <r_2, L\,|<r_1, R\,|\,H_{F'12}\,|\,r_2, L>|\,r_1, R> \qquad (22)$$

The non-correlated sum ultimately reduces to $4E_{11LL}$, because of $S_{iiXX} = 1$ and in view of the symmetry of indistinguishable electrons and indistinguishable vacancies. The correlated sum has a symmetry of its own leading to $4\varepsilon_{12LR}$, where

$$\varepsilon_{12LR} = <r_1, L | <r_2, R | H_{F'12} | r_2, L> | r_1, R>, \tag{23}$$

etc. We finally arrive at

$$E_{F'saddle} \approx 4 [E_{11LL} - \varepsilon_{12LR}]. \tag{24}$$

For a rough estimate of the correlated separation, we set $r_{12} \sim 2r_0$ (vacancy-centered configuration), $r_{12} \sim d_0 \sim r_0$ (saddle-point configuration) where $r_0$ is the cavity radius, $d_0$ is the semi-vacancy separation (from center to center along <110>). In an fcc alkali halide lattice $d_0 \sim \sqrt{2}a$ where $a$ is the anion-cation separation. It is also implied that the semi-vacancy radius is $\sim \frac{1}{2} r_0$, as opposed to the non-split vacancy radius $r_0$, twice as large.

The semi-vacancy energy $E_{11LL}$ may be inferred from Wang's data, though changing the effective charge from $-1$ to $-\frac{1}{2}$ and the cavity radius from $r_0$ to $\frac{1}{2} r_0$. Wang's data relating to the vacancy-centered configuration, we set $d_0 = 0$. In turn, the semi-vacancy pair situation obtains at finite $d_0$. It is the $\boldsymbol{d_0}$ vector associated with the halogen-ion mode along <110> which splits the full-vacancy V into two semi-vacancies at L and R. The full-vacancy hosts the entire F′ electron. Now the arguments leading to (19) above are in full force at L ≡ R ≡ V too and we get

$$E_{F'vacancy} = 8 [E_{11VV} - \tfrac{1}{2} \varepsilon_{12VV}] \tag{25}$$

for the vacancy-centered-configuration energy. The ½ factor to the correlation energy in (25) agrees with what has been said above about the average correlation length $r_{12} \sim r_0$ for a split vacancy and $r_{12} \sim\sim 2r_0$ for a non-split vacancy.

From (24) and (25) we derive the condition that the vacancy-centered F′ energy surpass the saddle-point F′ energy, $E_{F'vacancy} \geq E_{F'saddle}$, as $2E_{11VV} \geq E_{11LL}$, provided the correlation energies meet $\varepsilon_{12LR} \sim \varepsilon_{12VV}$, which is plausible in view of the above interpretation. Therein, we argued that $E_{11VV}$ and $E_{11LL}$ should be similar in magnitude.

## 5. Semicontinuum model

In so far as our primary objective has been illustrating the problem rather than solving it exactly, we will further on apply the "archaic" semicontinuum potentials to dealing with the F′ center problem. Moreover, the semicontinuum model is perhaps the simplest description of the saddle-point configuration involving the two semivacancies in Figure 1 and is also credible as a first order approximation to the point-ion potential.[1] Ultimately, this will enable one to derive easily eigenfunctions and eigenvalues for calculating $E_{11VV}$ and $E_{11LL}$ so as to obtain a quantitative criterion. The semicontinuum model being one of

a spherical well combined with a Coulomb tail outside, its radial bound state wave functions are:[9,10]

$\psi(r) = A_{Kl} j_l(Kr)$ (*in*)

$\psi(r) = B_{\alpha l} R(\alpha r)$ (*out*) (26)

where *in* and *out* stand for in-cavity at $r \leq r_0$ and out-of-cavity at $r \geq r_0$, $r_0$ is the cavity radius. Here $j_l(Kr)$ and $R(\alpha r)$ are the spherical Bessel functions and the hydrogen-like wave functions, respectively. The eigenvalues to (26) read

$E_\psi \equiv <\psi | H_{F'} |\psi> = [\eta^2 k^2/2m_e - V_0] <\psi|\psi>_{in} - [\eta^2 \alpha^2/2m_e] <\psi|\psi>_{out}$ (27)

where $V_0$ is the cavity potential, $m_e$ is the electron mass, and the remaining symbols have their usual meaning as in textbooks. The cavity radius and potential have three appearances below, namely as $r_{0i}$, and $V_{0i}$, respectively, with i = L, R, and V. Both cavity radius and potential can be dependent on a certain mode coordinate thereby effecting an electron-mode coupling. The eigenfunctions are normalized so that $<\psi|\psi>_{in}$ and $<\psi|\psi>_{out}$ only comprise respective parts of the electron clouds. The wave functions *in* and *out* should be continuous and derivative continuous at the $r = r_0$ boundary. This requirement secures two equations for $v = kr_0$ and $u = \alpha r_0$ which are to be solved. The two components of the continuous requirement are:

$A_{Kl} j_l(Kr_0) = B_{\alpha l} R(\alpha r_0)$

$A_{Kl} \, dj_l(Kr)/dr \big|_{r=r0} = B_{\alpha l} \, dR(\alpha r)/dr \big|_{r=r0}$ (28)

For example, the ground state semi-continuum radial wave-functions are:

$\psi(r) = A_{k0} (Kr)^{-1} \sin(Kr)$, *in*

$\psi(r) = B_{10} \exp(-\alpha r)$, *out* (29)

Setting $\alpha r_0 = u$ and $Kr_0 = v$, we get the following continuity equations:

$A_{k0} v^{-1} \sin(v) = B_{10} \exp(-u)$

$A_{k0} v [v \cos(v) - \sin(v)] / v^2 = -B_{10} u \exp(-u)$ (30)

Dividing up the two equations in (28) we find

$(1 - u) \sin(v) = v \cos(v)$, or $u = 1 - v \cotan(v)$ (31)

while from $<\psi(r)|\psi(r)>_{all\ space} = 1$ we get

$$A^2 r_0^3 / 2v^2 = \{1 - (1/v) \sin(v) \cos(v) + (1/u) \sin^2(v) [1 + (1/u) + (1/2u^2)]\}^{-1}$$

$$(B^2 r_0^3 / 2u) \exp(-2u) = [A^2 r_0^3 / 2v^2] \sin^2(v) / u \tag{32}$$

From the definition of a dielectric constant approppriate to the problem ($\varepsilon_\infty$ - optical dielectric constant, $\varepsilon_0$ - static dielectric constant)

$$\varepsilon^{-1} = \varepsilon_\infty^{-1} - (\varepsilon_\infty^{-1} - \varepsilon_0^{-1}) (B^2 r_0^3 / 2u) \exp(-2u)(1/8u) [2 + (1/u) + (3/4u^2)] / [1 + 1/2u] \tag{33}$$

From (33) and $\alpha = m_e e^2 / \varepsilon \eta^2$, $u = \alpha r_0$ we also get alternatively

$$u = u_\infty - (u_\infty - u_0) (B^2 r_0^3 / 2u) \exp(-2u) (1/8u) [2 + (1/u) + (3/4u^2)] / [1 + (1/2u)] \tag{34}$$

We also derive

$$<\psi_{1s} | \psi_{1s}>_{in} = (A^2 r_0^3 / 2v^2) [1 - (1/v) \sin(v) \cos(v)]$$

$$<\psi_{1s} | \psi_{1s}>_{out} = (B^2 r_0^3 / 2u) \exp(-2u) [1 + (1/u) + (1/2u^2)] \tag{35}$$

and the ground state eigenvalue reads

$$E_{1s} = [(\eta^2 v^2 / 2m_e r_0^2) - V_0] <\psi_{1s} | \psi_{1s}>_{in} - (\eta^2 u^2 / 2m_e r_0^2) <\psi_{1s} | \psi_{1s}>_{out} \tag{36}$$

with $<\psi_{1s} | \psi_{1s}>_{in}$ and $<\psi_{1s} | \psi_{1s}>_{out}$ from (34), as well as with a cavity potential

$$V_0 = (\alpha_M e^2 / r_0) - (e^2 / 2r_0) (1 - 1/\varepsilon_\infty) - \chi - (e^2 / 2r_0) (1/\varepsilon_\infty - 1/\varepsilon_0) \times$$

$$(B^2 r_0^3 / 2u) \exp(-2u) (1/u) [1 + (1/u)] \tag{37}$$

So far we have illustrated the method for ground state calculations mainly. Nevertheless, excited F′ states can be considered along similar lines. Indeed, further details of calculations carried out for 1s-, 2s-. and 2p- like states can be found elsewhere.[10]

The potential $V_0$ is most controversial. Apart from polarization effects, the potential "seen" by the F′ electron may be characteristic of a neutral center rather than of an attractive center, due to screening by the other electron. (This is an example where the simplifying assumption of an extra F′ electron may not work.) On the other hand, an attractive potential could arise in principle due to lattice relaxation effects (cf. ref. [3]). A alternative is provided by the semicontinuum approch to $V_0$ as above. We therefore are led to the conclusion that $V_0$ could be taken to mean an attractive potential whose nature is to be only specified independently.

The semicontinuum calculations are shown in Table I for a particular alkali halide (NaI). F′ centers in NaI have been found to exhibit peculiar optical properties,[11] such as the bell-shaped absorption band unlike the F′ bands in most "conventional" halides which are flat and structureless.[1] This behavior has stimulated our earlier research into whether the two extra electrons at the NaI F′ center are not bound by the negative-U mechanism in which lattice relaxation effects play the major role.[3] The result proved to be cautiously affirmative, since more data are needed to verify the hypothesis in the remaining alkali iodides.

Table I
Semicontinuum calculations

| Host NaI | State | $r_0$ (Å) | u | v | A(Å$^{-3/2}$) | B(Å$^{-3/2}$) | ε | $V_0$ (eV) | $-\varepsilon_\psi$ (eV) |
|---|---|---|---|---|---|---|---|---|---|
| fullvac | $\psi_{1s}$ | 3.237 | 0.98529 | 1.5615 | 0.20092 | 0.34464 | 3.10 | 5.00290 | 1.41496 |
| | $\psi_{2s}$ | | 0.51507 | 1.87 | 0.07221 | 0.12738 | 2.965 | 4.92352 | 0.24849 |
| | $\psi_{2p}$ | | 0.43510 | 1.4298 | 0.04150 | 0.05691 | 3.51 | 4.55741 | 0.08224 |
| semivac | $\psi_{1s}$ | 1.618 | 0.98529 | 1.5615 | 0.56829 | 0.97479 | 3.10 | 11.0058 | 3.08475 |
| | $\psi_{2s}$ | | 0.51507 | 1.87 | 0.20424 | 0.36028 | 2.965 | 10.8470 | 0.43917 |
| | $\psi_{2p}$ | | 0.43510 | 1.4298 | 0.11738 | 0.16097 | 3.51 | 10.1148 | 0.15015 |

A similar cautious statement can be made as to the feasibility of the saddle point conjecture to the F′ center in NaI. Indeed, from the last column in Table I we see that the ground-state F′ center (when *both* its electrons are in the $\psi_{1s}$ ground state) is energetically more stable at the full vacancy configuration than it is at the saddle point configuration. From the data therein we also see that the $\psi_{2s}$ and $\psi_{2p}$ electronic energies at the full- and semi- vacancy configurations (when both F′ electrons are in the excited electronic states) are more likely to meet the $2E_{11VV} \geq E_{11LL}$ criterion for the fullvac energy to surpass the semivac energy. We reserve some degree of uncertainty though, since the semicontinuum calculations are credible down to a certain extent so that more refined calculations will have to be planned.

6. Conclusion

We carried out an extensive investigation to determine whether the energy of the vacancy - centered configuration (VCC) can surpass the energy of the saddle-point configuration (SPC) for the F′ center in alkali halides. The study gives an affirmative result in some cases namely when the matrix element of the F′ Hamiltonian in VCC ground state exceeds half the size of the diagonal matrix element in SPC ground state. The case has been illustrated by semicontinuum calculations for F′ centers in NaI. They show that the F′ center might prefer the saddle point configuration as energetically more favorable if both its electrons are in the excited state. Formation of $^{2P}$F′ centers residing at saddle point could therefore be achieved through double excitation in a high-density photon field.

We accounted for the interaction of the electronic system with the < 110 > $B_{1u}$ phonon mode to incorporate the effect of a phonon coupling. This coupling splits the F′ vacancy into two semi-vacancies turning the F′ configuration molecule-like. In this respect the present study complements a previous analysis of the role of phonon coupling leading to estimates of the negative-U contribution to the F′ binding energy, as described in [3].

Details were also extended of a routine calculation of the SPC tunneling splitting originally from Ref. [5] to obtain a quantitative criterion for the distribution of the F′ density in and around the cavity.

Our conclusion that excited F′ centers in NaI may reside in energetically more favorable positions at saddle point sites rather than at full vacancy sites is similar to Luty's suggestion as regards the $F_A(II)$ centers in KCl:Li.[5] This suggestion has again been aimed at explaining the unusual optical behavior of type-II $F_A$ centers. We again refer the reader to Luty's excellent paper for greater details on the matter of saddle-point color centers..